\documentclass[conference]{IEEEtran}
\IEEEoverridecommandlockouts
\usepackage{cite}
\usepackage{amsmath,amssymb,amsfonts}
\usepackage{listings}
\usepackage{algorithmic}
\usepackage{graphicx}
\usepackage{textcomp}
\usepackage{xcolor}
\usepackage{url}
\usepackage{verbatim}
\usepackage{hyperref}
\def\BibTeX{{\rm B\kern-.05em{\sc i\kern-.025em b}\kern-.08em
    T\kern-.1667em\lower.7ex\hbox{E}\kern-.125emX}}
    
    \definecolor{lightGray}{gray}{0.9}
\begin{document}

\newcommand{\nb}[2]{
    \fbox{\bfseries\sffamily\scriptsize#1}
    {\sf\small\textcolor{red}{\textit{#2}}}
}
\newcommand\ag[1]{\nb{AG}{#1}}
\newcommand\tp[1]{\nb{TP}{#1}}
\newtheorem{puzzle}{Puzzle}

\title{Detecting diabetic retinopathy severity through fundus images using an ensemble of classifiers}

\author{
    \IEEEauthorblockN{Eduard Popescu\IEEEauthorrefmark{1}, Adrian Groza\IEEEauthorrefmark{1}, Ioana Damian\IEEEauthorrefmark{2}}
    \IEEEauthorblockA{\IEEEauthorrefmark{1}Department of Computer Science, \\
    Technical University of Cluj-Napoca, 400114 Cluj-Napoca, Romania
    \\\{Popescu.Pa.Eduard@student.utcluj.ro, adrian.groza@cs.utcluj.ro\}}
    \IEEEauthorblockA{\IEEEauthorrefmark{2}
Department of Ophthalmology, “Iuliu Hatieganu” University of Medicine and Pharmacy, \\
Emergency County Hospital, 400347 Cluj-Napoca, Romania
    \\\{inteioana@yahoo.com\}
}
}



\maketitle

\begin{abstract}
Diabetic retinopathy is an ocular condition that affects individuals with diabetes mellitus. 
It is a common complication of diabetes that can impact the eyes and lead to vision loss. 
One method for diagnosing diabetic retinopathy is the examination of the fundus of the eye. 
An ophthalmologist examines the back part of the eye, including the retina, optic nerve, and the blood vessels that supply the retina. 
In the case of diabetic retinopathy, the blood vessels in the retina deteriorate and can lead to bleeding, swelling, and other changes that affect vision.
We proposed a method for detecting diabetic diabetic severity levels. 
First, a set of data-prerpocessing is applied to available data: adaptive equalisation, color normalisation, Gausssian filter, removal of the optic disc and blood vessels. 
Second, we perform image segmentation for relevant markers and extract features from the fundus images.
Third, we apply an ensemble of classifiers and we assess the trust in the system.

\end{abstract}


\section{Motivation}
The number of people suffering from diabetes is increasing, with the majority of them facing diabetic retinopathy, a condition that requires specialized control by an ophthalmologist. 
Statistics have indicated that 80\% of patients with prolonged diabetes suffer from various stages of diabetic retinopathy.\cite{fong2004retinopathy}.
In many cases, persons either do not know they have this condition until it reaches an untreatable stage or they visit the ophthalmologist even when the condition is not yet present, leading to overcrowding in ophthalmology clinics and affecting patients who actually require specialized medical control at that time.

An early diagnostic process can reduce the development of diabetic retinopathy and the predisposition to severe blindness~\cite{neinih}. Considering that a medical consultation at an ophthalmology clinic to detect the stage or presence of diabetic retinopathy is a feasible but sluggish process, it is not recommended for early detection of diabetic retinopathy. It is essential to undergo this consultation as quickly as possible, preferably from the comfort of one's own home or any pharmacy, without the need for a doctor's visit.

We develop here a method for detecting of diabetic retinopathy (DR) from retinal images using various Machine Learning and Deep Learning techniques. The proposed method is implemented as a mobile application. 

\section{Diabetic retinopathy grading}

\subsection{Diabetic retinopathy severity levels}
Diabetic retinopathy (DR) is the leading cause of preventable vision impairment and blindness in the European Region. 
Approximately 1 in 3 people with diabetes mellitus (DM) will develop DR but its damaging effects on vision can be prevented by early detection and treatment through screening. 
Because 90 to 95\% of all patients with DM have type 2 DM (T2DM), a large proportion of patients with DR belong to this group, although type 1 DM (T1DM) is associated with more severe ocular complications~\cite{flaxel2020diabetic}. 

\begin{table*}
\caption{International clinical diabetic retinopathy disease severity scale. NPDR= nonproliferative diabetic retinopathy; IRMA= intraretinal microvascular abnormalities; PDR= proliferative diabetic retinopathy\label{tab:classification}}
\begin{tabular}{lp{14cm}}
Disease severity level & Findings observable upon dilated ophthalmoscopy \\ \hline
No apparent retinopathy & No abnormalities \\
Mild NPDR & Microaneurysms only\\
Moderate NPDR & More than just microaneurysms but less than severe NPDR\\
Severe NPDR & Any of the following and no signs of proliferative retinopathy:
More than 20 intraretinal hemorrhages in each of 4 quadrants
Definite venous beading in 2 or more quadrants
Prominent IRMA in 1 or more quadrants\\
PDR &  One or both of the following: Neovascularization or Vitreous/ preretinal hemorrhage
\end{tabular}
\end{table*}

Without an appropriate intervention, DR progresses from mild to more severe stages (See Table~\ref{tab:classification}). 
The nonproliferative DR (NPDR) stages are characterized by: 
(1) retinal vascular related abnormalities; 
(2) increased vascular permeability;
(3) impaired perfusion and retinal ischemia, if DR progresses and gradual closure of vessels appear.

Retinal vascular related abnormalities includes 
(i) microaneurysms: saccular localized outpouchings of the capillary wall seen as tiny red dots, indistinguishable clinically from dot hemorrhages.
(ii) intraretinal hemorrhages of three aspects: 
(a) retinal nerve fiber layer (RNFL) hemorrhages arise from the larger superficial precapillary arteriole, seen as flame shaped red lesions.
(b) intraretinal hemorrhages arise from the venous end of capillaries, with a configuration of red “dot/blot” lesion.
(c) deeper dark round hemorrhages are secondary to hemorrhagic retinal infarcts.
(iii) venous dilation 
(iv) cotton-wool spots: represent accumulation of neuronal debris within the RNFL. They appear as small fluffy whitish superficial lesions that can obscure the underlying blood vessels.

Increased vascular permeability translates into:
       (i) lipid deposits, seen as exudates: composed of lipoprotein and lipid-filled macrophages, seen as waxy yellow lesions with relatively distinct margins, localized in clumps and/or rings.
       (ii) retinal thickening, seen as edema. 

Impaired perfusion and retinal ischemia, if DR progresses and gradual closure of vessels appear, it would further translate into:
(i) venous abnormalities: dilation, beading, loops.
(ii) intra-retinal microvascular abnormalities (IRMA): arteriolar-venular shunts seen as fine, irregular and red intraretinal lines connecting arterioles to venules.
(iii) severe and extensive vascular leakage: increased retinal hemorrhages and exudation.

Clinically significant macular edema (CSME) represents retinal thickening and/or adjacent hard exudates that are either found in central subfield zone of 1 mm in diameter (center-involved diabetic macular edema (DME)) or outside of it (non-center involved DME).  CSME can occur at any stage of retinopathy: NPDR or PDR, and due to involvement of the central macula it threatens the vision.
Proliferative diabetic retinopathy (PDR) consists of neovascularization at the inner surface of the retina and also into the vitreous, secondary to more global retinal ischemia. New vessels on or near the optic disc (NVD) or new vessels elsewhere on the retina (NVE) could develop and if they rupture, vitreous hemorrhage (VH) or preretinal hemorrhage appear. 

The grading of DME is somehow problematic since stereo examination of the retina, which is suitable in detecting retinal thickening, could be achieved with either slit-lamp biomicroscopy or stereophotography. 
Because exudates are usually associated with significant macular edema, their presence might aid in suspecting or confirming DME (see Table~\ref{tab:grading}). Here RT stands for retinal thickening and PP for posterior pole.

\begin{table}
\caption{Diabetic macular edema disease severity scale\label{tab:grading}}
\begin{tabular}{ll}
Proposed DME severity level & Findings from dilated ophthalmoscopy \\ \hline
DME apparently absent     & No apparent RT or exudates in the PP\\
DME apparently present     & Some apparent RT or exudates in PP
\end{tabular}
\end{table}

The purpose of DR screening is to reduce the risk of vision impairment among asymptomatic patients with DM through the early identification of retinal changes and implementation of an effective treatment. American Academy of Ophthalmology recommends the first retinal examination for type 1 DM patients to be 3-4 years after DM diagnosis, while for type 2DM patients at the time of DM diagnosis. 

In 2018 International Council of Ophthalmology (ICO) and American Diabetes Association (ADA) released a guideline for DR screening, stating that an adequate protocol consists of visual acuity exam+ retinal examination through ophthalmoscopy (direct/indirect) or slit-lamp biomicroscopy or 30º to wide field retinal (fundus) monophotography or stereophotography, and dilated or undilated photography 
The retinal examination could be performed by trained personnel with no medical degree.  The guideline varies by country and availability of resources. 
   
Dilated fundus exam performed by an ophthalmologist is still considered the gold standard method for diagnosing DR. However, due to the high prevalence of DM patients and the limited number of ophthalmologists, this could prove unfeasible. Thus, the screening programs are transitioning to retinal photography-based screening which seems to be also cost effective.  Stereoscopic 35-mm retinal photography through dilated pupil with a specially trained retinal grader interpreting the image is the gold standard for retinal imaging. But the cost of the technique, interpretation and acquisition limits the application as a general screening method. Non-mydriatic retinal photography demonstrated similar sensitivity and specificity (78-98\%, 86-90\%, respectively) to a dilated fundus exam performed by an ophthalmologist (84-92\%, 92-98\%)~\cite{ting2016diabetic}. Tele-retina, consisting of retinal images acquired at one site, which are transmitted and interpreted at another site could be another cost-effective approach. 
       
For an accurate grading of DR, good quality images are essential. 
Technical errors were defined by Agrawal et al. as~\cite{agrawal2003technical}:
(i) photographic error such as image not well centered or poor clarity obscuring view of 1/3 or more of the temporal image or the large temporal blood vessels.
(ii) technical failure due to patient factors such as: media opacity (cataract, corneal opacities, vitreous opacities), small pupil, patients with difficulty in positioning.
The proportion of ungradable images seems to be greater for nonmydriatic than mydriatic examinations: 30\% versus 10\% ungradable, as shown by Kim et al.~\cite{kim2007accuracy}.

\subsection{Applying AI in DR screening}
DR detection could be enhanced by artificial intelligence (AI) methods by reducing the burden of manual review of fundus photography and also reducing the need for trained graders thus enabling a more efficient screening. A systematic review from 2023 which evaluated the diagnostic value of AI algorithm models for DR found a pooled sensitivity of 0.880 (0.875-0.884), a pooled specificity 0.912 (0.99-0.913), a pooled positive likelihood ratio 3.021 (10.738-15.789), a pooled negative likelihood ratio 0.083 (0.061-0.112), area under the curve 0.9798 and a pooled diagnostic odds ratio 206.80 (124.82-342.63)~\cite{wang2023performance}.

Predictive modeling is useful in identifying patients at risk of DR progression and further defining personalized screening intervals. 
IDx-DR (Digital Diagnostics, Coralville, IA, USA), started as a machine learning (ML) algorithm under the name of Iowa detection program (IDP), which achieved a sensitivity of 96.8\% and specificity of 69.4\% in detecting referable DR (rDR) on the Messidor-2 dataset 6. The later version of IDP, named IDx-DR was a combination of convolutional neural networks (CNN) and deep learning (DL) enhancement.  IDx-DR X2.1, which is a newer version enhanced by DL components, achieved a sensitivity of 96.8\% and a specificity of 87\% for rDR, and a sensitivity of 100\% and a specificity of 90.8\% for vision-threatening DR (VTDR)~\cite{abramoff2016improved}. 
It is the first authorized AI device to detect DR, and was approved by FDA in 2018. This particular software utilizes for analysis two 45-degree photographs, one disc-center and another one macula-center, acquired with a non-mydriatic, non-ultrawide field camera. The results provided for the doctor are either 1)” more than mild diabetic retinopathy detected: refer to an eye care professional” or 2)” negative for more than mild diabetic retinopathy: rescreen in 12 months”. 

EyeART (Eyenuk Inc., Los Angeles, CA, USA) started as a ML-based algorithm, and was validated on 5084 DM patients from EyePACS and on 40.542 images from another EyePACS dataset, proving a sensitivity of 90.0\% and a specificity of 63.2\%~\cite{bhaskaranand2016automated}. 
The newer version is EyeArt v2.1, DL-based version, which achieved a higher sensitivity (91.3\%) and a higher specificity (91.1\%) after analyzing $>$800.000 patients from DR screening protocol of EyePACS9. In august 2020, FDA approved EyeArt v2.2.0 to identify more than mild and VTDR, using Canon CR-2AF and Canon CR-2 Plus AF cameras, while in June 2023 received clearance to use the Topcon NW400 retinal camera, thus becoming the first and only AI system that is FDA-cleared to be used with multiple retinal cameras by different manufacturers. 
In the European Union, it is the first and only AI system approved under MDR Class IIb to detect DR, age-related macular degeneration, and glaucomatous optic nerve damage, in a single test. In a pivotal, prospective, multicenter clinical trial from 2021, EyeART demonstrated high sensitivity 96\% for more than mild DR and 97\% for VTDR, and also high specificity 88\% for more than mild DR and 90\% for VTDR10.

\section{Detecting diabetic rethinophaty with ensemble learning}
The system consists of two modules. 
The first module is designed for early classification, where classification is performed based on extracted features from the image using a hybrid method similar to the one described above. 
The second module contains an ensemble of U-Net networks, each specializing in the segmentation of a specific type of lesion.

\subsection{Dataset used}
Two datasets were used. 
The first one is the "Asia Pacific Tele-Ophthalmology Society (APTOS)" dataset~\cite{aptos_dataset}, which aids in the classification of retinal images. 
The second one is the "Indian Diabetic Retinopathy Image Dataset (IDRiD)"~\cite{h25w98-18}, used for diabetic retinopathy lesion segmentation.
The first dataset is used for training and validating the convolutional networks for extracting features from retinal images. 
The dataset contains images of various sizes depicting different stages of diabetic retinopathy. 
The second dataset is used for the segmentation algorithm, which will train the U-Net network. It consists of 54 training images and 27 test and validation images. For each image, there are four masks representing the location of each lesion, along with masks for segmenting the optic disc.


\begin{figure*}
    \centering
    \includegraphics[width=0.2\textwidth, keepaspectratio]{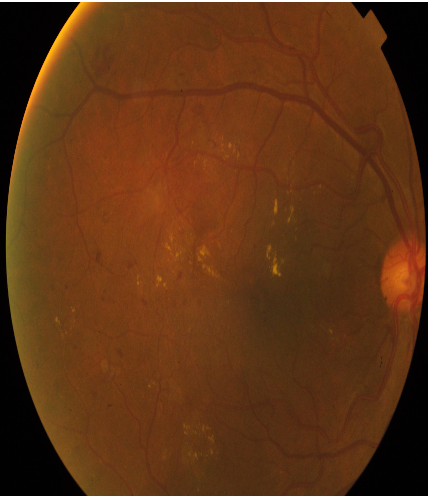} 
    \hfill
    \includegraphics[width=0.2\textwidth, keepaspectratio]{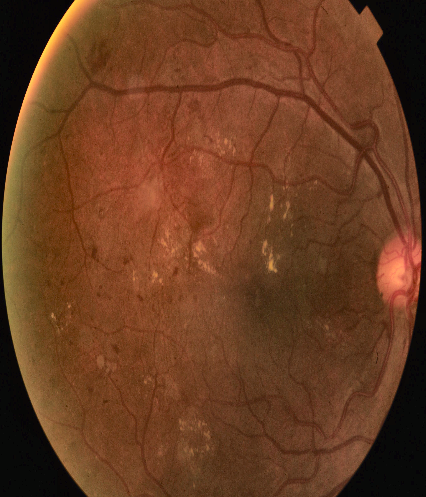}
    \hfill
    \includegraphics[width=0.2\textwidth, height=0.2\textheight, keepaspectratio]{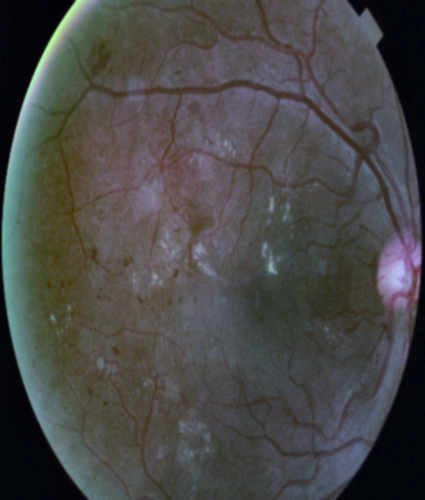}
    \hfill 
    \includegraphics[width=0.3\textwidth]{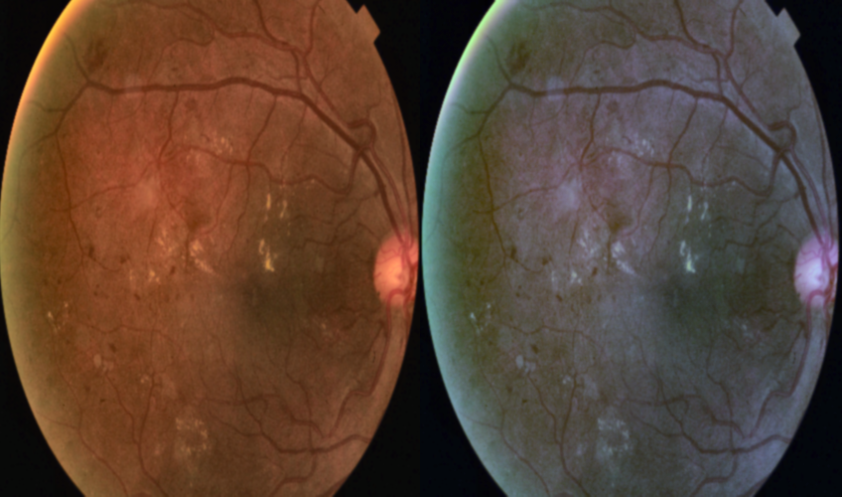} 
    \caption{Data preprocessing: original message (1$^{st}$), after adaptive histogram equalization (2$^{nd}$), after color normalisation (3$^{rd}$), 
    after gaussina filter (4$^{th}$) }
    \label{fig:claheimg}
\end{figure*}

\subsection{Data preprocessing} The following preprocessing steps have been performed:
\paragraph{Adaptive equalization of histograms in the RGB model}
Adaptive histogram equalization enhances the quality of images, emphasizes important details, and ensures consistency and comparability among the available images. 
By enhancing contrast and highlighting details, adaptive histogram equalization can aid in more precise detection and classification of objects or regions of interest. 
After the adaptive equalization in Figure~\ref{fig:claheimg}, one can see that .?????....

\paragraph{Color normalization}
Some classification models can be sensitive to color variations and may produce inconsistent results depending on the predominant colors in the image. 
By normalizing colors, we can minimize this sensitivity and achieve more robust and stable classification. 
Images can exhibit variations in contrast and brightness, which can affect interpretation and feature extraction. 
By normalizing colors, we can balance the contrast and brightness of the images, facilitating accurate detection and classification of objects.

\paragraph{Gaussian filter}
A Gaussian filter is constructed to remove noise from the image. 
The filter parameters are $\sigma_x$, $\sigma_y$, $\mu_x$ and $\mu_y$. 
The Gaussian kernel is calculated using these parameters, and the kernel is normalized. The Gaussian filter is applied to the CLAHE image. The result is a filtered image with the noise removal effect.

\begin{figure}
    \centering
    \includegraphics[width=0.35\textwidth, height=0.4\textheight, keepaspectratio]{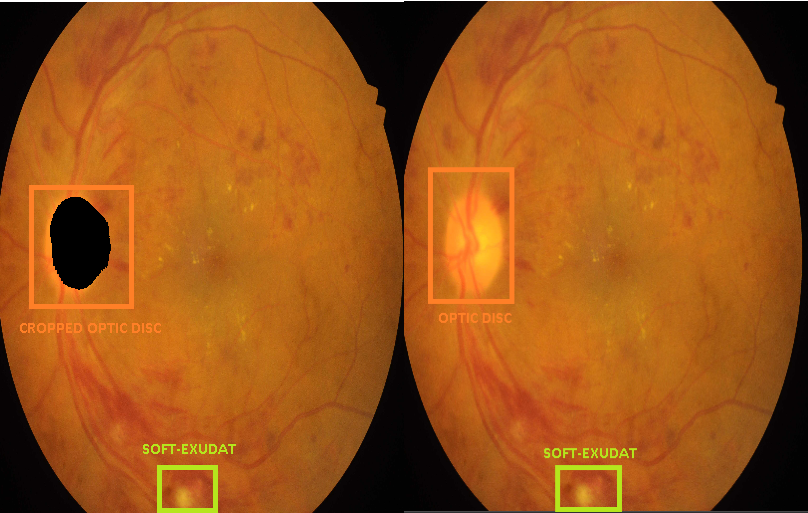}
    \caption{Image of the fundus after and before removing the optic disc}
    \label{fig:ODImg}
\end{figure}

\paragraph{Removal of the optic disc} 
Since the eye fundus image contains the optic disc, whose edges are diffuse and can be confused with the edges of a soft exudate, we opt to remove the optic disc (see Figure~\ref{fig:ODImg}).

\begin{figure}
    \centering
    \includegraphics[width=0.35\textwidth]{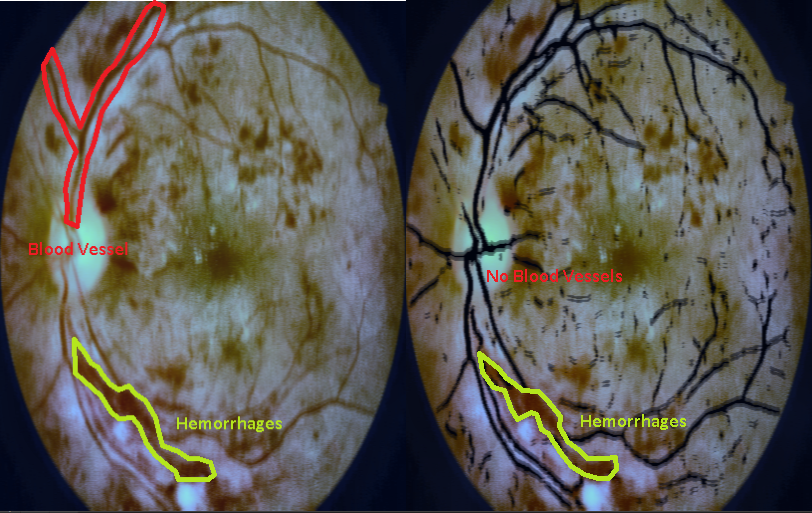}
    \caption{From left to right the initial image and the image without blood vessels result}
    \label{fig:BV_CROOP}
\end{figure}

\paragraph{Removing and blood vessels} 
Blood vessels can be easily confused by the model with hemorrhages. 
The main preprocessing operation for these lesions is the extraction of blood vessels from the image, thereby eliminating the possibility of segmenting a blood vessel as a hemorrhage or microaneurysm (see Figure~\ref{fig:BV_CROOP}).

The optic disc and blood vessels segmentation will be performed using a trained U-Net model to identify the optic disc and blood vessels in a retinal fundus image. Once the mask indicating the location, the disc and vessels can be accurately cropped from the image.

\subsection{Hybrid feature extraction and classification module}
We analysed the perfomance of the available models for the early diagnostic phase. 

\begin{figure}
    \centering
    \includegraphics[width=0.3\textwidth]{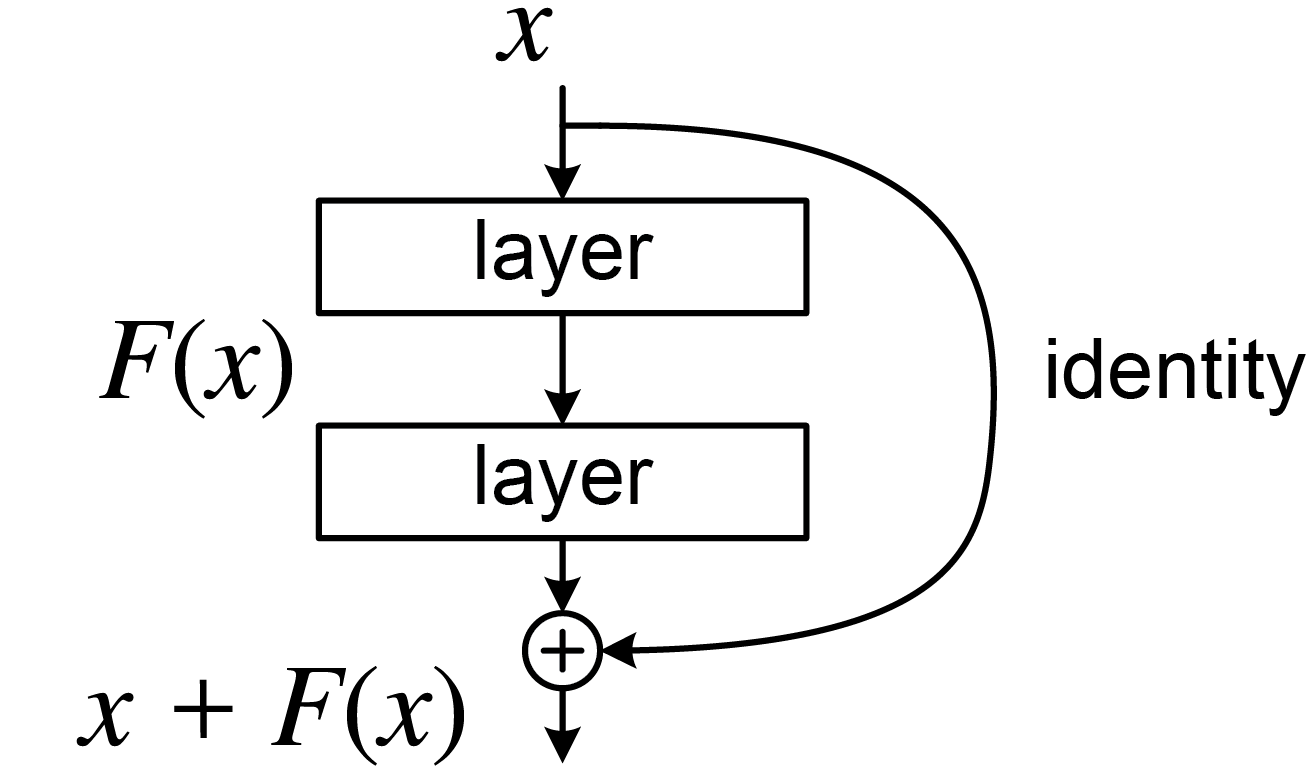}
    \caption{Residual connections between two layers}
    \label{fig:resnet}
\end{figure}

The ResNet model, known for its extensive use in image classification, particularly in classifying fundus images, was one option. 
Performance analysis indicates that ResNet50 yields the best results within the ResNet family. 
However, due to memory constraints and hardware considerations in the context of a mobile application, ResNet18 is chosen as a suitable alternative, offering comparable performance with reduced resource requirements.
ResNet18 has 18 layers, including convolutional, pooling, and fully connected layers. 
It consists of an input convolutional layer followed by four main blocks, each containing multiple convolutional layers. 
The blocks utilize residual connections by adding the input layer's results to their output, creating shortcuts that enable direct passage of information as in Figure~\ref{fig:resnet}.

The  convolutional layers that play a crucial role in extracting features. 
The convolution is performed with the equation (\ref{eq:conv}):
\begin{equation}
g(x, y) = \sum_{i=-a}^{a} \sum_{j=-b}^{b} f(x+i, y+j) \times k(i, j) \label{eq:conv}    
\end{equation}

We also examined GoogleNet (or Inception V1), which demonstrates impressive feature extraction capabilities for fundus images in our experiments. The GoogleNet architecture stands out for its use of an innovative convolutional module called the "Inception module", which helps reduce the number of parameters and the complexity of the network.
The network architecture described includes an important block called the Auxiliary block. This block consists of convolutional and pooling layers, followed by a fully connected layer and a softmax layer. Its purpose is to provide a secondary path for gradient propagation during training and offer additional information for image classification. 
The Auxiliary block is specific to the GoogleNet architecture and is used during training to improve learning and achieve better performance in image classification. During training, backpropagation is applied to the losses calculated in this block, facilitating efficient information propagation and enhancing the overall model performance.
Although Inception V3 performs notably well in classifying diabetic retinopathy, its larger dimensions and resource demands are deemed less favorable compared to Inception V1. As a compromise, a prioritization is made for higher accuracy, resulting in increased time and memory usage.

The VGG-19 and DenseNet-201 models exhibit impressive prediction quality but are costly to implement in parallel due to their high number of trainable parameters.
Overall, the analysis considers the performance and efficiency trade-offs of various deep learning models, ultimately selecting ResNet18 and Inception V1 as suitable choices for the proposed system (see Table~\ref{tab:models}).
The results from the table \ref{tab:models} also suggest overfitting and an imbalance in the data distribution. However, the analyses conducted have revealed that these values are due to an imbalance in the data distribution. Therefore, it was decided to reduce the number of classes from five to three since the dataset suffers from the problem of imbalance, namely without DR (stage 0), mild DR (stage 1-2), and severe DR (stage 3-4).

\begin{table}
        \centering
        \label{tab:models}
\caption{Analysing the available models for feature extraction}
\begin{tabular}{lccr}
        Model & Training accuracy & Validation accuracy & Parameters  \\
        \hline
        ResNet50    & 99.37   & 71.64  &  24$\times 10^6$   \\
        ResNet18    & 93.50   & 69.34  & 11$\times 10^6$    \\
        VGG-19    & 97.98   & 73.37  & 138$\times 10^6$    \\
        DensNet-201    & 99.58   & 76.80  & 21$\times 10^6$   \\
        GoogleNet    & 96.58   & 74.34  & 6$\times 10^6$    \\
        InceptionV3  & 99.03   & 73.72  & 24$\times 10^6$   \\
\end{tabular}
\end{table}

The final component of the hybrid feature extraction and classification module is the block of automatic learning algorithms, which will receive 1056 features from both ResNet18 and GoogleNet networks and classify the retinal fundus images into the following categories: no-DR, mild-DR, severe-DR. This decision will be made based on the weighted voting technique where each machine learning algorithm will have a weight depending on its performance on the validation set.\\
We reduce the number of classes from five to three because the dataset suffers from the imbalance problem, that is, no DR (stage 0), mild DR (stage 1-2), and severe DR. (stage 3-4).

\subsection{Segmentation ensemble of U-Net networks}
The U-Net architecture features a symmetric design consisting of an encoding part and a decoding part, which communicate through a channel at the central level of the architecture.
The encoding part consists of consecutive convolutional layers followed by activation functions and pooling layers, which reduce the spatial dimensions of the extracted features. This section is responsible for capturing contextual information and reducing the spatial size.
The decoding part consists of transposed convolutional layers that increase the spatial dimensions of the extracted features. During the expansion process, features from deeper levels of the encoder are concatenated with the features from the expansion part, helping preserve contextual information.

U-Net generates a binary image that represents the probability of each pixel belonging to the desired type of lesion. Starting from this probability image, a threshold is chosen to construct a mask containing only the lesions with the highest probability. Finally, after the lesions have been identified and their number determined, a decision is made regarding the stage of diabetic retinopathy exhibited in the input image.

\section{Results}
First, we analyse the impact of preprocessing steps.
Second, we assess the performance of the lession segmentation.

\subsection{The impact of preprocessing} 

\begin{table}
\centering
\caption{Performance of models with and without preprocessing}
\label{tab:Performancetable}

\begin{tabular}{|cllll|}
\hline
\multicolumn{5}{|c|}{\textbf{Train}}                                                                                                                                                \\ \hline
\multicolumn{1}{|l|}{Model} & \multicolumn{2}{l|}{Images with preprocessing}                   & \multicolumn{2}{l|}{Images without preprocessing}                 \\ \cline{2-5} 
\multicolumn{1}{|c|}{}                       & \multicolumn{1}{c|}{Accuracy} & \multicolumn{1}{c|}{Loss} & \multicolumn{1}{c|}{Accuracy} & \multicolumn{1}{c|}{Loss} \\ \hline
\multicolumn{1}{|l|}{ResNet18}               & \multicolumn{1}{l|}{98.41}         & \multicolumn{1}{l|}{0.052}        & \multicolumn{1}{l|}{94.54}         & 0.125                            \\ \hline
\multicolumn{1}{|l|}{GoogleNet}              & \multicolumn{1}{l|}{97.67}         & \multicolumn{1}{l|}{0.061}        & \multicolumn{1}{l|}{95.61}         & 0.110                             \\ \hline
\multicolumn{5}{|c|}{\textbf{Validation}}                                                                                                                                                 \\ \hline
\multicolumn{1}{|l|}{Model} & \multicolumn{2}{l|}{Images with preprocessing}                   & \multicolumn{2}{l|}{Images without preprocessing}                 \\ \cline{2-5} 
\multicolumn{1}{|c|}{}                       & \multicolumn{1}{c|}{Accuracy} & \multicolumn{1}{c|}{Loss} & \multicolumn{1}{c|}{Accuracy} & \multicolumn{1}{c|}{Loss} \\ \hline
\multicolumn{1}{|l|}{ResNet18}               & \multicolumn{1}{l|}{97.09}         & \multicolumn{1}{l|}{0.025}        & \multicolumn{1}{l|}{54.37}         & 1.933                             \\ \hline
\multicolumn{1}{|l|}{GoogleNet}              & \multicolumn{1}{l|}{96.91}         & \multicolumn{1}{l|}{0.030}        & \multicolumn{1}{l|}{60.72}         & 1.425                             \\ \hline
\end{tabular}
\end{table}

\begin{figure}
    \centering
    \includegraphics[width=0.5\textwidth]{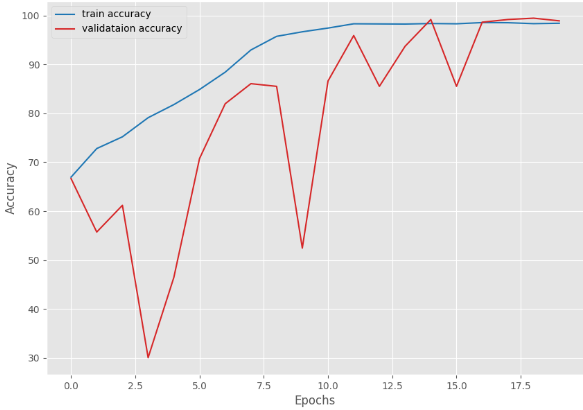}
    \caption{Improving the accuracy of ResNet18}
    \label{fig:resnetAcc}
\end{figure}

\begin{figure}
    \centering
    \includegraphics[width=0.5\textwidth]{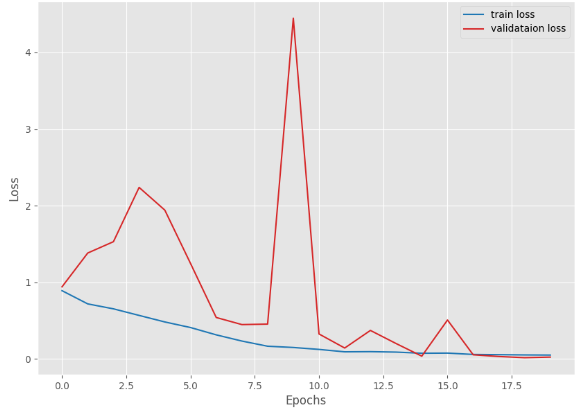}
    \caption{ResNet18 network loss mitigation}
    \label{fig:resnetLoss}
\end{figure}

\begin{table}
\centering
\caption{Classification accuracy according to the stage of DR}
\label{tab:mlTable}
\begin{tabular}{lccc}
Model                                                             & No-DR & Mild-DR & Server-DR \\ \hline
SVM Linear Kernel       & 96.90 & 90.15   & 71.15     \\ 
SVM Polynomial Kernel   & 95.34 & 89.01   & 75.95     \\ 
SVM Radial Basis Kernel & 96.71 & 90.90   & 66.92     \\ 
SVM Cramér-Singer       & 96.43 & 89.39   & 73.74     \\ 
Random Forest                                              & 92.87 & 86.74   & 53.73     \\ 
Naive Bayes                                                   & 79.45 & 83.71   & 71.42     \\ 
\end{tabular}

\end{table}
Both training and validation were conducted with a batch size of 32. The training process consisted of 20 epochs, starting with a learning rate of 0.01. Validation was performed after each epoch to observe the progression of the models. Such progression can be visualized in Figures \ref{fig:resnetAcc} and \ref{fig:resnetLoss}, which demonstrate the improvement of the ResNet18 model from one epoch to another using the preprocessed dataset.

Table~\ref{tab:mlTable} presents the performance of machine learning models on each type of classification. It can be observed that certain models perform differently on certain classification classes. Therefore, to provide the final result, it was decided not to rely on a single model for classification, but to use all models as an ensemble, where the decision is made based on a technique called "weight voting".

\subsection{The performance of the lesion segmentation module}
\begin{figure}
    \centering
    \includegraphics[width=0.4\textwidth, height=0.4\textheight, keepaspectratio]{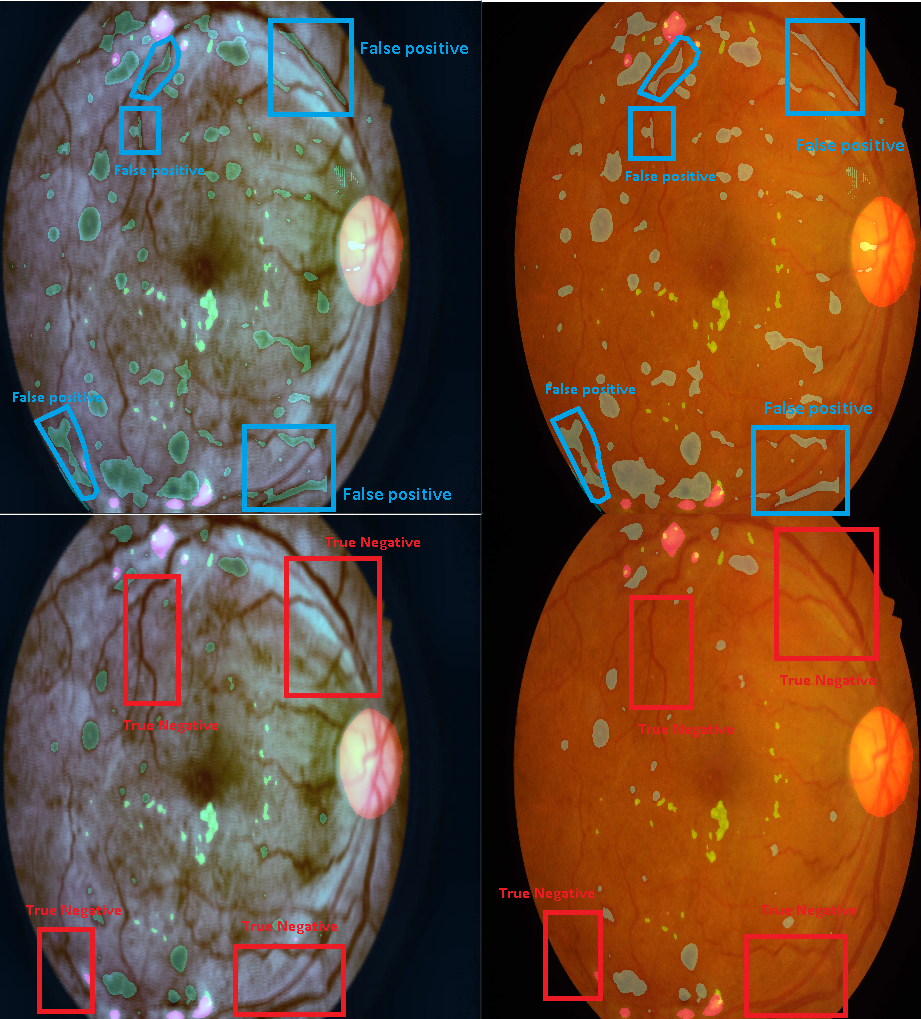}
    \caption{Segmentation result with and without blood vessels (dark green areas)}
    \label{fig:fp1fp2}
\end{figure}

Figure~\ref{fig:fp1fp2} illustrates that the removal of blood vessels in the process of segmenting microaneurysms and hemorrhages leads to a reduction in the number of false positives.

\begin{figure}
    \centering
    \includegraphics[width=0.4\textwidth]{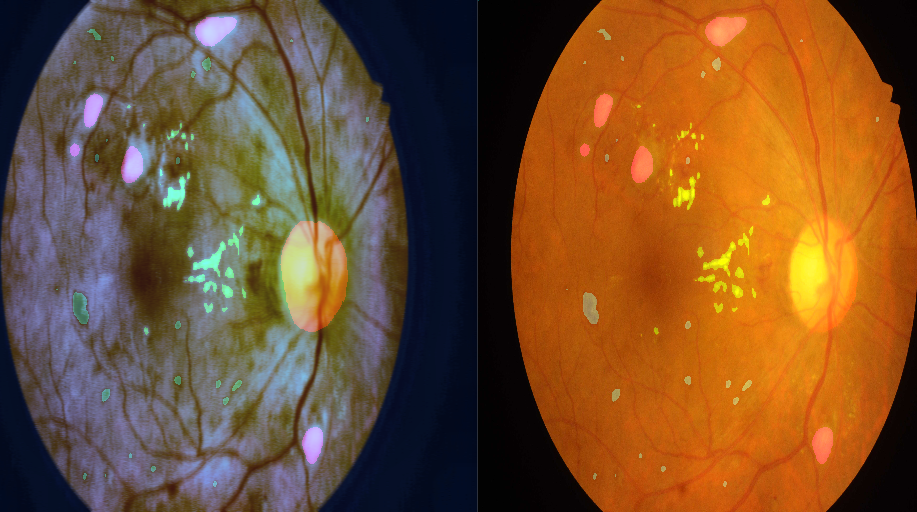}
    \caption{Segmentation result of soft exudates (areas in pink)}
    \label{fig:se_out}
\end{figure}

Next, in Figure~\ref{fig:se_out}, the result of segmenting soft exudates is presented. 
It can be observed that in this case, due to the absence of the optic disc and the applied filter, the model manages to achieve accurate segmentation despite the limited training data.

\begin{figure}
    \centering
    \includegraphics[width=0.4\textwidth]{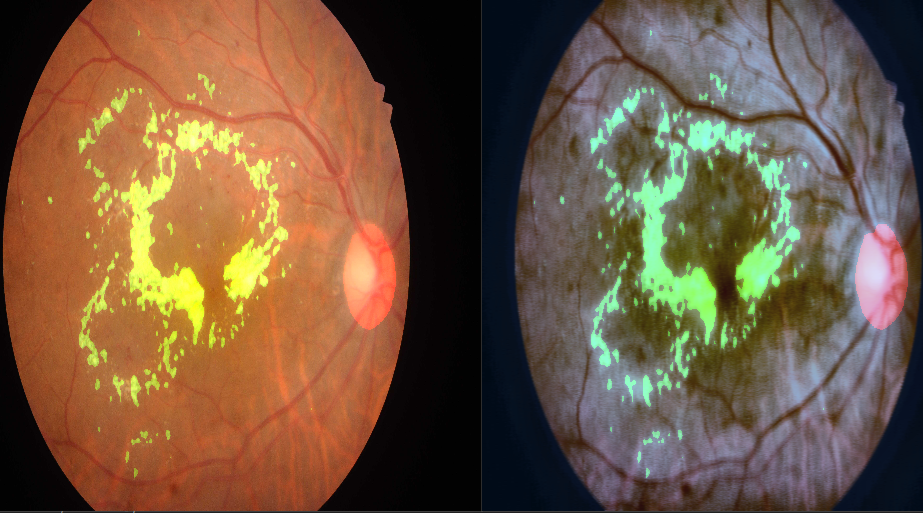}
    \caption{Segmentation result of hard exudates (neon green areas)}
    \label{fig:ex_out}
\end{figure}

Figure~\ref{fig:ex_out} shows the result of segmenting hard exudates. 
The U-Net model achieved remarkable results in identifying this type of lesion due to the large amount of training data and the clarity of the lesions in the image. 
Additional preprocessing was not needed for detecting these lesions, only the initial preprocessing, which can be visualized on the right side of the figure.

\begin{figure}
    \centering
    \includegraphics[width=0.4\textwidth, height=0.4\textheight, keepaspectratio]{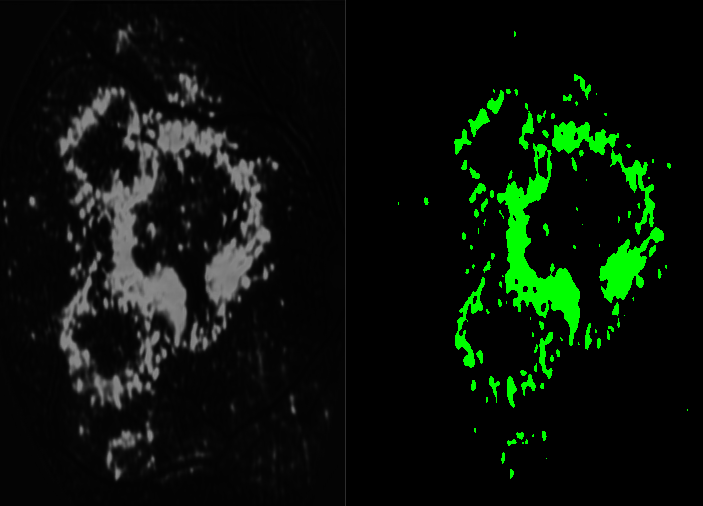}
    \caption{The probability mask (left) and the mask generated by the Otsu method (right)}
    \label{fig:ex_out_pm}
\end{figure}

The confidence of the model in segmenting soft exudates can be interpreted by looking at Figure~\ref{fig:ex_out_pm}, where the probability mask representing the probability matrix is shown. The content of the mask suggests that the model assigns high probabilities to pixels within the lesion and a probability close to 0 to pixels outside the lesion. Therefore, the probability mask closely aligns with the applied mask shown in Figure \ref{fig:ex_out}. This indicates that the model exhibits low uncertainty compared to previous models.
\subsection{Explainable AI in the segmentation module}
For each segmentation model, a confidence score is calculated, represented as a percentage. 
The formula used to calculate it is a weighted average between the quantified image quality, the F1 score, and the model's confidence in segmenting the respective lesion.

\begin{table}
\centering
\caption{The weights corresponding to each metric}
\label{tab:ponders}
\begin{tabular}{|c|c|c|c|}
\hline
Metrics & Image Quality Score & F1 Score & Confidence Score \\ \hline
Weights & 0.4                                                           & 0.3                                                & 0.3                                                        \\ \hline
\end{tabular}
\end{table}
The weights for each metric (Table \ref{tab:ponders}) are calculated based on experiments that observed the influence on the segmentation model's performance. 
Images with poor quality tend to be incorrectly segmented due to the loss of information caused by the lack of clarity.

To calculate the confidence with which the model indicates that a pixel belongs to a lesion, the entropy of the probability distribution is calculated with $Entropy = \sum_{i=1}^{n} p_i \log(p_i)$, 
where $n$ is the number of distinct events in the distribution, while 
$p_i$ is the probability of pixel $i$ belonging to the lesion.
The entropy is computed for each input image. 
In the end, the result will be $1$ minus the mean entropy (Table~\ref{tab:metrics}).

To calculate the final confidence (last line in Table~\ref{tab:metrics}), 
which represents the segmentation module's confidence for each lesion, the quantification of input image quality is also required. 
This is done using the "Mean Squared Error (MSE)" formula, which measures the average squared difference between two images. 
It is used in this case to evaluate the quality between a reference image representing the image with the best quality found up to that point and the image provided to the system.
This metric assesses the trust in the module and tis value is shown to the user for awareness of limitations in Fig.~\ref{fig:se_out}.
\begin{table}
\centering
\caption{Confidence and IoU values for each lesion in the Figure~\ref{fig:se_out}}
\label{tab:metrics}
\begin{tabular}{lllll}
Lesion &  HEM & SE & HE & MA \\ \hline
Confidence & 0.94 & 0.89& 0.91 & 0.60 \\
IoU score & 0.85 & 0.92 & 0.97 & 0.77 \\ \hline 
Module Trust & 87\%  & 90\% & 93\% & 76\%\\ 
\end{tabular}
\end{table}

For the segmentation dataset, considering that the lesions are identified by an external person and there is no certainty that all identified lesions are accurate, the Cohen's Kappa coefficient was calculated to measure the agreement or reliability among two evaluators statistically.
$K = \frac{(P_o - P_e)}{(1 - P_e)}$, 
where $K$ represents Cohen's Kappa coefficient, $P_o$ represents the observed proportion of interrater agreement, and 
  $P_e$ represents the proportion of agreement expected by chance.
The interpretation of this coefficient is: 
\begin{center}
    \begin{tabular}{ll}
  $K < 0$ &  The agreement is weaker than by chance\\
  $K = 0$ &  Agreement equals chance. \\
  $0 < K < 0.2$ &  Weak agreement.\\
  $0.2\leq K < 0.4$ &  Moderate agreement.\\
  $0.4 \leq K < 0.6$ &  Medium agreement.\\
  $0.6 \leq K < 0.8$ &  Significant agreement.\\
  $0.8 \leq K \le 1$ &  Almost perfect match. 
\end{tabular}
\end{center}
In our case, this agreement was 0.73 which means that we have a significant agreement.

\section{Related Work}
\begin{figure}
    \centering
    \includegraphics[width=0.4\textwidth, height=0.4\textheight, keepaspectratio]{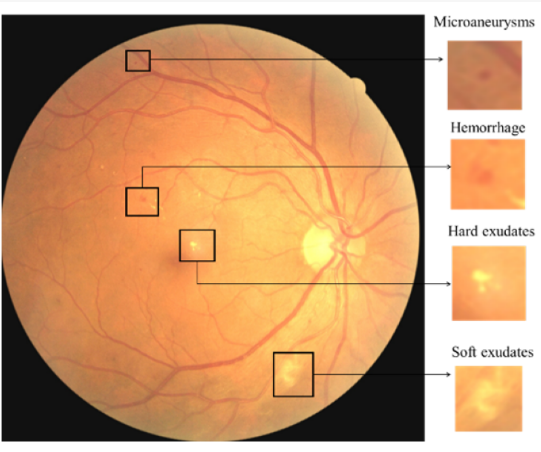}
    \caption{Different lesions in a fundus image~\cite{s21113704}}
    \label{fig:wwdr}
\end{figure}

Diagnosis is done through the analysis of retinal images extracted from the patient, images that require collection using state-of-the-art ophthalmic devices such as OCT~\cite{muntean2023predictive},~\cite{marginean2022low} or fundus images~\cite{butt2022diabetic},~\cite{bdcc6040146},~\cite{Marupally2017}. 
By analyzing these images, various lesions can be extracted, as shown in Figure~\ref{fig:wwdr}, indicating the stage of DR.
Butt et al.~\cite{butt2022diabetic} have proposed a hybrid learning method based on extracting features from retinal images using two specialized convolutional neural networks in the field of image recognition: GoogleNet and ResNet18.
The two convolutional neural networks (CNNs) combine the features extracted from the image, specifically 1.000 from GoogleNet and 1.000 from ResNet18, resulting in a total of 2.000 features for the image to be classified. 
These features feed multiple automated learning algorithms such as Support Vector Machine, Random Forest, Radial Basis Function, and Naïve Bayes, with the target to classify the stage of diabetic retinopathy.

A hybrid approach is also proposed by Novitasari et al.~\cite{bdcc6040146}.
The method is based on feature extraction using one of the convolutional neural networks mentioned earlier, along with DenseNet, which is specialized in image recognition similar to GoogleNet and ResNet18. 
The classifier is a Deep Extreme Learning Machine. 
The training method using the DELM algorithm provides better prediction than that of convolutional neural networks.

On the line of segmenting lesions in retinal images, Marupally et al.~\cite{Marupally2017} have work on identifying hard exudates with different types of exposure, both well-defined and diffuse ones.
The RGB image is transformed into grayscale for better exposure of the bright areas that may contain hard exudates. 
The grayscale image is then passed through a top-hat filter with a structural element radius of $\frac{1}{4}$ of the image width, followed by histogram equalization and binarization with a threshold of 0.37.
To identify diffuse hard exudates, the first phase involves extracting the green plane from the image, resulting in a grayscale image. The next step, similar to the well-defined lesions, is passing it through a top-hat filter with a structural element radius of $\frac{1}{20}$ of the image width. 
Finally, histogram equalization is performed, followed by binarization with a threshold of 0.25.

A segmentation model for identifying and delineating lesions in various medical conditions is the U-Net architecture~\cite{ronneberger2015u}. 
U-Net is recognized for its innovative approach of combining local detail information with global contextual information, enabling precise image segmentation. 
The U-Net architecture consists of an encoder and a decoder, which are connected through a skip connection. 
This allows the transfer of information from the encoder to the decoder while preserving higher-resolution details during processing.
Relying on segmentation instead of black box machine learning, one can take decisions similar to rule-based medica protocols~\cite{sluavescu2014towards}.

Eftekhari et al.~\cite{Eftekhari2019} 
have focused on identification of microaneurysms, which are defining lesions for certain stages of DR. 
These biomarkers are extremely difficult to identify due to their small size and uneven distribution. 
A CNN is trained with patches of size 101$\times$101 from the image to create a probability map where the intensity of each pixel represents the probability of that pixel belonging to a microaneurysm. 
Then, patches that do not contain a microaneurysm are eliminated based on the probabilities previously identified. 
The remaining patches are provided to another specialized CNN, which generates a mask indicating the identified microaneurysms.


\section{Conclusion}
We proposed a method for detecting severity levels of the diabetic retinopathy condition. 
By leveraging deep learning techniques and utilizing a carefully curated dataset, the model achieves high accuracy in distinguishing different stages of the disease. 
The use of image preprocessing techniques, such as adaptive histogram equalization and vessel segmentation, enhances the robustness and reliability of the classification system. 
The ensemble of multiple models further improves the overall performance by leveraging the strengths of each individual model. The results highlight the potential of the developed approach for accurate and efficient diabetic retinopathy classification, which could aid in early diagnosis and timely treatment. 
This research contributes to the advancement of computer-aided diagnosis systems for diabetic retinopathy, with the potential to improve patient care and reduce the burden on healthcare professionals.
\bibliographystyle{IEEEtran} 
\bibliography{bib}

\begin{thebibliography}{10}
\providecommand{\url}[1]{#1}
\csname url@samestyle\endcsname
\providecommand{\newblock}{\relax}
\providecommand{\bibinfo}[2]{#2}
\providecommand{\BIBentrySTDinterwordspacing}{\spaceskip=0pt\relax}
\providecommand{\BIBentryALTinterwordstretchfactor}{4}
\providecommand{\BIBentryALTinterwordspacing}{\spaceskip=\fontdimen2\font plus
\BIBentryALTinterwordstretchfactor\fontdimen3\font minus
  \fontdimen4\font\relax}
\providecommand{\BIBforeignlanguage}[2]{{%
\expandafter\ifx\csname l@#1\endcsname\relax
\typeout{** WARNING: IEEEtran.bst: No hyphenation pattern has been}%
\typeout{** loaded for the language `#1'. Using the pattern for}%
\typeout{** the default language instead.}%
\else
\language=\csname l@#1\endcsname
\fi
#2}}
\providecommand{\BIBdecl}{\relax}
\BIBdecl

\bibitem{fong2004retinopathy}
D.~S. Fong, L.~Aiello, T.~W. Gardner, G.~L. King, G.~Blankenship, J.~D.
  Cavallerano, F.~L. Ferris~III, R.~Klein, and A.~D. Association, ``Retinopathy
  in diabetes,'' \emph{Diabetes care}, vol.~27, no. suppl\_1, p. 84 87, 2004.

\bibitem{neinih}
\BIBentryALTinterwordspacing
``Diabetic retinopathy, updated (2022),'' Jul. 2022. [Online]. Available:
  \url{https://www.nei.nih.gov/learn-about-eye-health/eye-conditions-and-diseases/diabetic-retinopathy}
\BIBentrySTDinterwordspacing

\bibitem{flaxel2020diabetic}
C.~J. Flaxel, R.~A. Adelman, S.~T. Bailey, A.~Fawzi, J.~I. Lim, G.~A.
  Vemulakonda, and G.-s. Ying, ``Diabetic retinopathy preferred practice
  pattern{\textregistered},'' \emph{Ophthalmology}, vol. 127, no.~1, pp.
  P66--P145, 2020.

\bibitem{ting2016diabetic}
D.~S.~W. Ting, G.~C.~M. Cheung, and T.~Y. Wong, ``Diabetic retinopathy: global
  prevalence, major risk factors, screening practices and public health
  challenges: a review,'' \emph{Clinical \& experimental ophthalmology},
  vol.~44, no.~4, pp. 260--277, 2016.

\bibitem{agrawal2003technical}
A.~Agrawal and M.~A. McKibbin, ``Technical failure in photographic screening
  for diabetic retinopathy,'' \emph{Diabetic medicine}, vol.~20, no.~9, pp.
  777--777, 2003.

\bibitem{kim2007accuracy}
H.~M. Kim, J.~C. Lowery, and R.~Kurtz, ``Accuracy of digital images for
  assessing diabetic retinopathy,'' \emph{Journal of Diabetes Science and
  Technology}, vol.~1, no.~4, pp. 531--539, 2007.

\bibitem{wang2023performance}
Z.~Wang, Z.~Li, K.~Li, S.~Mu, X.~Zhou, and Y.~Di, ``Performance of artificial
  intelligence in diabetic retinopathy screening: a systematic review and
  meta-analysis of prospective studies,'' \emph{Frontiers in Endocrinology},
  vol.~14, p. 1197783, 2023.

\bibitem{abramoff2016improved}
M.~D. Abr{\`a}moff, Y.~Lou, A.~Erginay, W.~Clarida, R.~Amelon, J.~C. Folk, and
  M.~Niemeijer, ``Improved automated detection of diabetic retinopathy on a
  publicly available dataset through integration of deep learning,''
  \emph{Investigative ophthalmology \& visual science}, vol.~57, no.~13, pp.
  5200--5206, 2016.

\bibitem{bhaskaranand2016automated}
M.~Bhaskaranand, C.~Ramachandra, S.~Bhat, J.~Cuadros, M.~G. Nittala, S.~Sadda,
  and K.~Solanki, ``Automated diabetic retinopathy screening and monitoring
  using retinal fundus image analysis,'' \emph{Journal of diabetes science and
  technology}, vol.~10, no.~2, pp. 254--261, 2016.

\bibitem{aptos_dataset}
\BIBentryALTinterwordspacing
{Herrero, Mar{\'i}a}, ``Diabetic retinopathy detection aptos dataset,'' Kaggle
  Datasets, 2019, accessed on 12 March 2023. [Online]. Available:
  \url{https://www.kaggle.com/datasets/mariaherrerot/aptos2019}
\BIBentrySTDinterwordspacing

\bibitem{h25w98-18}
\BIBentryALTinterwordspacing
P.~Porwal, S.~Pachade, R.~Kamble, M.~Kokare, G.~Deshmukh, V.~Sahasrabuddhe, and
  F.~Meriaudeau, ``Indian diabetic retinopathy image dataset (idrid),'' 2018.
  [Online]. Available: \url{https://dx.doi.org/10.21227/H25W98}
\BIBentrySTDinterwordspacing

\bibitem{muntean2023predictive}
G.~A. Muntean, A.~Marginean, A.~Groza, I.~Damian, S.~A. Roman, M.~C. Hapca,
  M.~V. Muntean, and S.~D. Nicoar{\u{a}}, ``The predictive capabilities of
  artificial intelligence-based oct analysis for age-related macular
  degeneration progression—a systematic review,'' \emph{Diagnostics},
  vol.~13, no.~14, p. 2464, 2023.

\bibitem{marginean2022low}
A.~Marginean, V.~Bianca, S.~D. Nicoara, and G.~Muntean, ``Low-dimensional
  representation of oct volumes with supervised contrastive learning,'' in
  \emph{2022 IEEE 18th International Conference on Intelligent Computer
  Communication and Processing (ICCP)}.\hskip 1em plus 0.5em minus 0.4em\relax
  IEEE, 2022, pp. 47--54.

\bibitem{butt2022diabetic}
M.~M. Butt, D.~N. F.~A. Iskandar, S.~E. Abdelhamid, G.~Latif, and R.~Alghazo,
  ``Diabetic retinopathy detection from fundus images of the eye using hybrid
  deep learning features,'' \emph{Diagnostics (Basel)}, vol.~12, no.~7, p.
  1607, Jul 2022.

\bibitem{bdcc6040146}
\BIBentryALTinterwordspacing
D.~C.~R. Novitasari, F.~Fatmawati, R.~Hendradi, H.~Rohayani, R.~Nariswari,
  A.~Arnita, M.~I. Hadi, R.~A. Saputra, and A.~Primadewi, ``Image fundus
  classification system for diabetic retinopathy stage detection using hybrid
  cnn-delm,'' \emph{Big Data and Cognitive Computing}, vol.~6, no.~4, 2022.
  [Online]. Available: \url{https://www.mdpi.com/2504-2289/6/4/146}
\BIBentrySTDinterwordspacing

\bibitem{Marupally2017}
\BIBentryALTinterwordspacing
A.~G. Marupally, K.~K. Vupparaboina, H.~K. Peguda, A.~Richhariya, S.~Jana, and
  J.~Chhablani, ``Semi-automated quantification of hard exudates in colour
  fundus photographs diagnosed with diabetic retinopathy,'' \emph{BMC
  Ophthalmology}, vol.~17, no.~1, p. 172, 2017. [Online]. Available:
  \url{https://doi.org/10.1186/s12886-017-0563-7}
\BIBentrySTDinterwordspacing

\bibitem{s21113704}
\BIBentryALTinterwordspacing
W.~L. Alyoubi, M.~F. Abulkhair, and W.~M. Shalash, ``Diabetic retinopathy
  fundus image classification and lesions localization system using deep
  learning,'' \emph{Sensors}, vol.~21, no.~11, 2021. [Online]. Available:
  \url{https://www.mdpi.com/1424-8220/21/11/3704}
\BIBentrySTDinterwordspacing

\bibitem{ronneberger2015u}
O.~Ronneberger, P.~Fischer, and T.~Brox, ``U-net: Convolutional networks for
  biomedical image segmentation,'' in \emph{International Conference on Medical
  image computing and computer-assisted intervention}.\hskip 1em plus 0.5em
  minus 0.4em\relax Springer, 2015, pp. 234--241.

\bibitem{sluavescu2014towards}
R.~R. Sl{\u{a}}vescu, A.-C. Gro{\c{s}}an, and K.~C. Sl{\u{a}}vescu, ``Towards
  assisting medical decisions by using rule based protocols and semantic
  resources,'' in \emph{International Conference on Advancements of Medicine
  and Health Care through Technology; 5th--7th June 2014, Cluj-Napoca, Romania:
  MEDITECH 2014}.\hskip 1em plus 0.5em minus 0.4em\relax Springer, 2014, pp.
  31--36.

\bibitem{Eftekhari2019}
\BIBentryALTinterwordspacing
N.~Eftekhari, H.~R. Pourreza, M.~Masoudi, K.~Ghiasi-Shirazi, and E.~Saeedi,
  ``Microaneurysm detection in fundus images using a two-step convolutional
  neural network,'' \emph{Biomedical Engineering Online}, vol.~18, no.~1,
  p.~67, May 2019. [Online]. Available:
  \url{https://doi.org/10.1186/s12938-019-0675-9}
\BIBentrySTDinterwordspacing

\end{thebibliography}

\end{document}